\documentclass[sigconf]{acmart}
\usepackage{csquotes}

\renewcommand\footnotetextcopyrightpermission[1]{}
\setcopyright{none}
\acmConference[LOCO 2026]{2nd International Workshop on Low Carbon Computing}{10--11 September 2026}{Lancaster, UK}

\usepackage{xcolor}
\usepackage{graphicx}
\usepackage{xurl}

\title{Turning interest into institutional change: teaching advocacy for sustainable research}

\author{Kirsty Pringle}
\affiliation{%
  \institution{EPCC, University of Edinburgh}
  \country{UK}
}
\email{k.pringle@epcc.ed.ac.uk}

\author{Lorna Smith}
\affiliation{%
  \institution{EPCC, University of Edinburgh}
  \country{UK}
}

\author{Erinma Ochu}
\affiliation{%
  \institution{University of West of England}
  \country{UK}
}

\author{Greg Wilson}
\affiliation{%
\institution{Independent Consultant}
\country{Canada}
}

\begin{document}

\begin{abstract}
Systemic change across the digital research landscape is required to reduce the environmental impact of digital research, but while many researchers and technical professionals are motivated to act, they often lack the skills required to translate motivation into lasting organisational change. We present an open-access course that teaches the foundations of advocacy and organisational change to researchers, research software engineers, and research technical professionals \citep{pringle2026advocacy}. Structured around the UNICEF five-step advocacy cycle, the course covers stakeholder analysis, power mapping, coalition building, storytelling, framing and messaging, and evaluation. It is grounded in the UK policy landscape, including the Concordat for Environmental Sustainability and the UKRI Environmental Sustainability Strategy, and uses a fictional case study to make concepts concrete. The course is designed as a dual-layer resource in which  workshop slides and extended self-study notes are contained in a single source material. A short pilot was delivered at the NetDRIVE Net Zero DRI Summer School in June 2026. The course is freely available under a CC BY 4.0 licence.  
\end{abstract}

\maketitle

\section{Introduction}

Environmental movements have often focused on raising awareness, and for low carbon computing this is an  important step because the environmental impact of ICT is largely invisible to the user. Unfortunately, however, increased awareness alone does not automatically translate into reduction in environmental impact.  Reduction can only be achieved by systemic change across the entire landscape, something that can be achieved either by \enquote{top down} interventions such as policies, incentives or legislation or by \enquote{bottom up} grass roots change. Most often, change occurs as a result of a combination of both these approaches.  The \enquote{bottom up}, grass roots change often relies on motivated individuals who essentially campaign for change, and while there are many individuals and emerging communities working towards changes that will reduce the environmental impact of ICT, there is little available training to support them in creating the change they are working towards. As a result of this, many sustainability interventions fail because while they raise awareness, they do not create institutional change. 

To address this, we have designed a training course that teaches the foundations of advocacy and organisational change, with the aim of empowering these motivated individuals, to help them create large scale change. The course is primarily designed for research computing audience, though the concepts can be applied to other sectors. Part of the motivation of targeting digital research is that in the research sector, there are already a number of high level (\enquote{top down}) strategies that aim to achieve change, but among the wider research community there is relatively little awareness of their content, or how they can be used to create  change.  

The course is open access and freely available, it is designed to be delivered as a 1-day workshop, though it comes with extensive notes and background information, making it accessible without attending a training event \citep{pringle2026advocacy}.

\section{Background and Motivation}
\label{background}

\begin{figure*}[htbp]
\includegraphics[width=\textwidth]{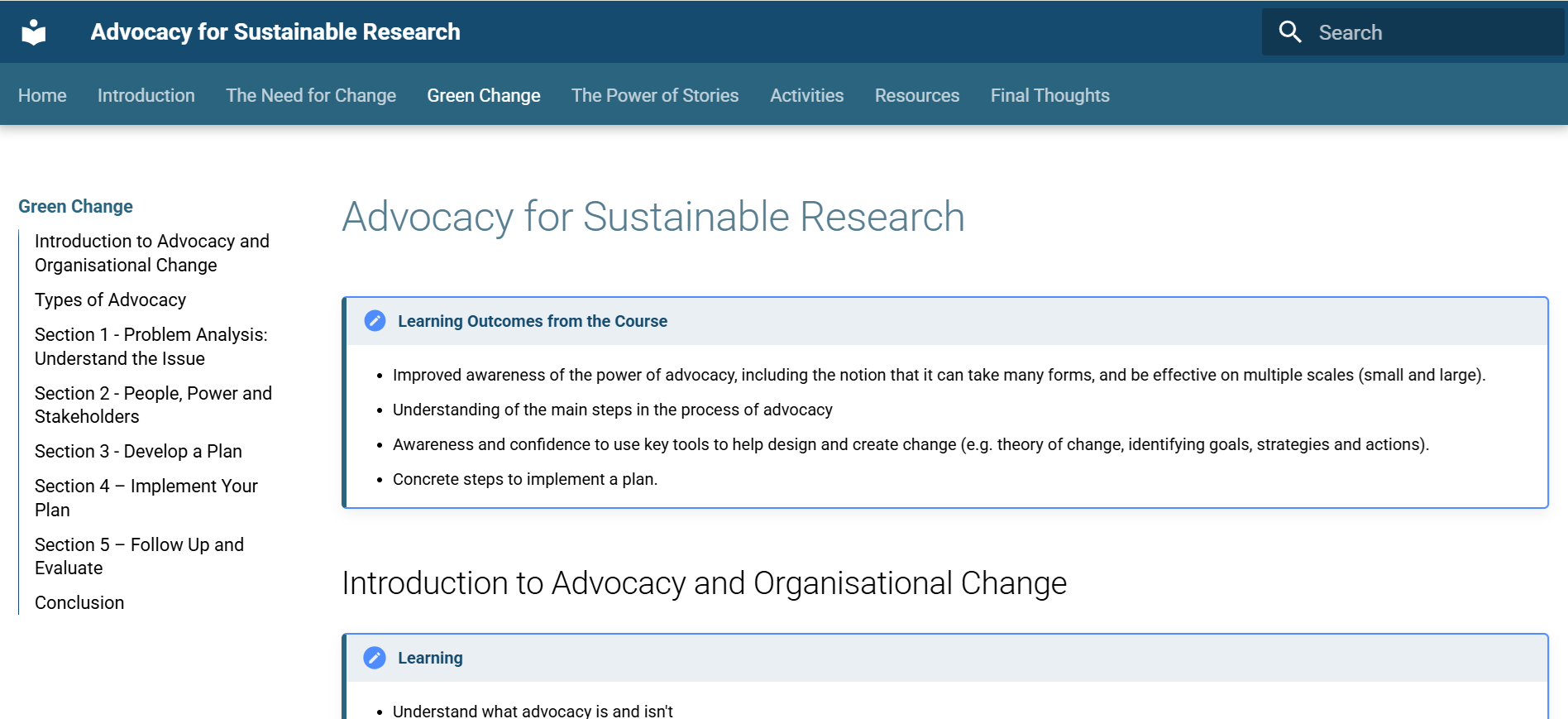}
\caption{Screenshot from the Advocacy for Sustainable Research course page showing the dual-layer format.}
\label{fig:screenshot}
\end{figure*}

The barrier to environmental change in research institutions is often not due to a lack of interest in or awareness of the issue.  For example, in  a recent survey by the University of Oxford of staff and students found very high levels of concern about environmental issues, with an average score of 4.5 out of 5 \citep{oxford2026}, but they also noted a \enquote{clear gap between grass-roots efforts and visible, strategic, institution-led change}.  Similarly, \citet{dablander2024} investigated scientists’ engagement with climate change using a mix of quantitative and qualitative analyses  and found that \enquote{Many scientists already engage in individual lifestyle changes, but fewer engage in advocacy or activism}.  This suggests that while awareness and motivation are often present, researchers may require additional skills and support to translate concern into organisational change. 

In addition to this staff and student engagement, environmental sustainability is increasingly embedded within UK research policy. The \textit{Concordat for the Environmental Sustainability of Research and Innovation Practice} commits signatories to integrating sustainability into research activities and reporting progress against their commitments \citep{concordat2024}. Similarly, the \textit{UKRI Environmental Sustainability Strategy 2025--2030} establishes expectations around leadership, infrastructure, procurement, travel and environmental reporting across the research sector \citep{ukri2025}. In addition to these sector-wide initiatives, most research organisations now maintain their own environmental sustainability strategies or net-zero plans. These policies provide a strong mandate for change, but awareness of them is often limited outside specialist sustainability teams. This policy landscape creates opportunities for advocacy. Rather than arguing for entirely new objectives, advocates can often use existing institutional and sector-wide commitments as leverage to support organisational change, but staff are unlikely to have had support or training to help them to \textit{use} these strategies.

In the past decade, significant improvements in wet lab environmental sustainability have been driven by technicians using certification schemes such as LEAF (\url{https://www.ucl.ac.uk/sustainable/leaf}) and My Green Lab (\url{https://mygreenlab.org}), which provide structured guidance and community support. More recently, the Green DiSC certification for digital research is now also gaining traction (\url{https://www.software.ac.uk/GreenDiSC}). The UKRI Net Zero Digital Research Infrastructure Scoping Project \citep{juckes2023} found that what was often lacking in the digital research community was not knowledge, but coordination and knowledge sharing. The report suggested that \enquote{green} research software engineers (RSEs) could play a similar role to wet lab technicians in driving change. A Green RSE Special Interest Group has been established to support this (\url{https://socrse.github.io/green-sig/}), but RSE remains a relatively small community. Reaching the broader population of digital research technical professionals including data scientists, research infrastructure engineers, data stewards, and digital researchers is likely to be necessary for change at scale.

The digital research landscape is complex and diverse, making it challenging to identify training needs across the community. The \enquote{Greening Digital Research} project led by Weronika Filinger (EPCC) made an important start, mapping the green skills required to help research institutions meet their environmental sustainability targets. The study was carried out in collaboration with the NetDRIVE (\url{https://uknetdrive.org}), DisCouRSE (\url{https://discourse-network.github.io}) and CHARTED (\url{https://drtp-skills.ac.uk/about/charted/})
Network+ projects, and used workshop activities and interviews with digital research technical professionals and researchers working with digital methods and tools. Advocacy was identified as both a key skill for driving environmental sustainability in research, and one for which no suitable training currently exists. This gap provided the direct motivation for the course described in this paper.

\begin{table}[h]
\small
\begin{tabular}{p{3.2cm}p{4.3cm}}
\toprule
\textbf{Section} & \textbf{Description} \\
\midrule
\textbf{Target Audience} & Early career researchers, research software engineers, research technical professionals, sustainability ambassadors \\
\textbf{Course Aim} & Equip motivated researchers with the advocacy and organisational change skills needed to drive environmental sustainability within their institutions. \\
\midrule
\href{https://kirstypringle.github.io/advocacy-for-sustainable-research/}{Home} &
Landing page and overview of the course. \\
\href{https://kirstypringle.github.io/advocacy-for-sustainable-research/introduction/}{Introduction} &
Introduces the course, its aims, and how to use the material. \\
\href{https://kirstypringle.github.io/advocacy-for-sustainable-research/green-change/the-need-for-change/}{The Need for Change} &
Background on environmental change, emissions, and the environmental footprint of research, for participants who need foundational context (see Section~\ref{sec:need-for-change}). \\
\href{https://kirstypringle.github.io/advocacy-for-sustainable-research/green-change/green-change/}{Green Change} &
The core taught content, structured around the five-step advocacy cycle (see Table~\ref{tab:course-structure}). \\
\href{https://kirstypringle.github.io/advocacy-for-sustainable-research/green-change/the-power-of-stories/}{The Power of Stories} &
Explores storytelling as a tool for advocacy and organisational change (see Section~\ref{sec:storytelling}). \\
\href{https://kirstypringle.github.io/advocacy-for-sustainable-research/activities/activities-all/}{Activities} &
Practical exercises and activities linked to the Greendale University case study. \\
\href{https://kirstypringle.github.io/advocacy-for-sustainable-research/resources/resources/}{Resources} &
Further reading and supporting materials for self-directed study. \\
\href{https://kirstypringle.github.io/advocacy-for-sustainable-research/final-thoughts/}{Final Thoughts} &
Closing reflections and next steps for participants. \\
\bottomrule
\end{tabular}
\caption{Course overview, background, resources and taught material}
\label{tab:course-sections}
\end{table}
\section{Course Design} 

The course is based on material from a 1-day course on organisational change developed by Greg Wilson for the research computing community (\url{https://third-bit.com/change/}).  The original course focuses broadly on how change occurs within organisations, placing particular emphasis on the role individuals can play in initiating, influencing and sustaining change. This people-centred perspective provided a natural foundation for developing a sustainability-focused course.

The course is designed for early career researchers, research software engineers and research technical professionals who are motivated to drive environmental change within their institutions but have no prior training in advocacy or organisational change. Participants are assumed to be familiar with research environments and to have an existing interest in sustainability, but no specialist knowledge is required. The course is explicitly designed for people acting in an unpaid, informal capacity - sustainability ambassadors rather than dedicated sustainability professionals.

In addition to the main taught material (Green Change), the course has additional information including background sections, activities and additional resources (see Table~\ref{tab:course-sections}).

\subsection{Green Change}
The main part of the advocacy course is within the \enquote{Green Change} section. It is structured around the UNICEF five-step advocacy cycle: (1) understanding the issue, (2) stakeholder analysis, (3) developing a plan, (4) implementation, and (5) evaluation and follow-up; a summary is given in Table~\ref{tab:course-structure}. See Figure~\ref{fig:screenshot} for a screenshot from the course.

\begin{table}[h]
\small
\begin{tabular}{p{3.2cm}p{4.3cm}}
\toprule
\textbf{Step} & \textbf{Content} \\
\midrule
1. Understand the Issue &
Policy landscape; Concordat for Environmental Sustainability; UKRI Environmental Sustainability Strategy; institutional commitments as advocacy leverage \\
2. People, Power and Stakeholders &
Power mapping; selectorate theory; stakeholder analysis; personas \\
3. Develop a Plan &
Goals, strategies and tactics; framing and messaging; theory of change \\
4. Implement Your Plan &
Diffusion of innovations; coalition building; navigating institutional politics; changing minds and handling resistance \\
5. Follow Up and Evaluate &
Outputs, outcomes and impact; following up on commitments; evaluation and iteration \\
\bottomrule
\end{tabular}
\caption{Structure and content of the Green Change taught section, built around the five-step advocacy cycle}
\label{tab:course-structure}
\end{table}
Other advocacy and change models could also have been used as a basic structure. Kotter's change model \citep{kotter1996leading} takes an 8-step approach: (1) create urgency, (2) build a guiding coalition, (3) form a vision, (4) communicate the vision, (5) remove obstacles, (6) generate short-term wins, (7) build on the change, (8) anchor changes in the culture. Alternatively, some domains use theory-of-change frameworks, which map desired outcomes to actions in a structured way. While theory-of-change frameworks are very useful, they were judged to be too complex to teach well in a short workshop. The UNICEF advocacy cycle covers similar ground to Kotter's eight-step model, but consolidates it into five steps, and fitted well with the structure of the original organisational change course on which this material is built. It was also selected because it presents change as an iterative process rather than a linear checklist, encouraging participants to revisit and refine their approach as circumstances change.

Similar to the source course material, there is a focus on practical application over theory. Throughout the course, concepts are illustrated through a fictional but realistic case study, Greendale University, which participants revisit across multiple steps of the advocacy cycle. The scenario includes a high-performance computing service and a cast of characters with competing priorities designed to reflect the realities of decision making within research organisations.

The course begins by introducing the policy landscape, including the Concordat for the Environmental Sustainability of Research and Innovation Practice, and explores how existing institutional commitments can be used as leverage for change. Subsequent steps cover stakeholder analysis, power mapping, selectorate theory and personas; goal setting, framing and theory of change; and approaches to implementation including diffusion of innovations, coalition building and navigating institutional politics. The final step focuses on evaluation, distinguishing between outputs, outcomes and longer-term impact. 

\subsection{The Need for Change}
\label{sec:need-for-change}

It is largely assumed that course participants will have an interest in, and basic foundational knowledge of environmental change thus there is relatively little content on this in the taught course.  However, to cover     any gaps there is also a \enquote{\href{https://kirstypringle.github.io/advocacy-for-sustainable-research/green-change/the-need-for-change/}{The Need for Change}} section which provides  background material on environmental change, Scope 1,2, and 3 emissions and the environmental footprint of research. The new training on \enquote{Emissions of Digital Research Infrastructure} on the SparkHUB platform is recommended pre-course work (\url{https://sdratraining.sparkhub.eu/}).

\subsection{Storytelling as an Advocacy Tool}
\label{sec:storytelling}
Storytelling can be an effective advocacy tool. A well-framed story can help people make sense of complex issues, and can shape decisions in ways that dry facts and figures often can't \citep{buganza2023storymaking, parkermoon2023digital, denning2000springboard, snow2021storytelling}.

Leadership stories, like \enquote{burning platform} narratives about why the status quo can't continue, or \enquote{vision} stories about a better future, help make the case for change and get people pulling in the same direction. Storytelling also works bottom-up, through personal stories that connect the everyday reality of a job to the bigger picture of institutional change -- and this local practice can, in turn, spread to wider systems change, as accounts of what worked in one institution inform advocacy elsewhere.

The topic is covered in the course in three ways: a background section introducing the concept, integration throughout the Green Change material, and use in the example exercises. The background section, \enquote{\href{https://kirstypringle.github.io/advocacy-for-sustainable-research/green-change/the-power-of-stories/}{The Power of Stories}}, introduces four story types (burning platform, vision, springboard, and champion or trailblazer) and describes how storytelling operates both top-down and bottom-up. Similar to The Need for Change section, this is designed to be an optional additional introduction section.

The four story types are also referenced within the taught sections of the advocacy cycle, covered in the \enquote{\href{https://kirstypringle.github.io/advocacy-for-sustainable-research/green-change/green-change/\#introduction-to-advocacy-and-organisational-change}{Green Change}} course:

\begin{itemize}
\item \textbf{Step 2 of the Green Change cycle (People, Power and Stakeholders)} -- each persona a participant writes is paired with the story type judged most likely to be effective for that stakeholder.
\item \textbf{Step 3 of the Green Change cycle (Develop a Plan)} -- storytelling is presented as one of three advocacy approaches, alongside data-driven and policy-based advocacy.
\item \textbf{Step 4 of the Green Change cycle (Implement Your Plan)} -- storytelling appears again in relation to coalition-building, where a first-hand account from a credible external voice is presented as an effective way to build support.
\end{itemize}

The persona-writing exercise in Step 2 asks participants to select a stakeholder from the Greendale case study and identify which story type would be most effective for them. The evaluation exercise in Step 5 asks participants to collect stories from their own follow-up work, for potential use in future advocacy.

Stories can also be  an inclusive way to create change; they don't require formal authority - someone with a compelling first-hand account can shift a conversation just as effectively as someone with positional power. This makes storytelling a particularly useful tool for the unpaid, informal advocates this course is aimed at, giving them a way to influence decisions even without the platform a senior leader has.

\subsection{Dual Layer Design}
The course is designed as a dual-layer resource: the same markdown source generates both workshop slides for facilitated delivery and extended web-based notes for self-directed learning. This allows the material to be delivered as a one-day workshop or studied independently. The course is published as an open-access resource using MkDocs and Material for MkDocs on GitHub Pages. 

It is intended to complement existing sustainability training resources such as Green DiSC, SparkHUB's Sustainable Digital Research training, and the EMBL-EBI Green Computing course by focusing specifically on advocacy and organisational change.

\section{Pilot and Evaluation Framework}
A short version of the course was piloted with approximately 25 PhD students at the NetDRIVE Net Zero DRI summer school in Durham (\url{https://durham.readthedocs.io/en/latest/netdrive/ss2026.html}) in June 2026. Informal feedback from the students was very positive; in a \enquote{show of hands} exercise, only 4 of the students had previously heard of the Concordat. One student commented afterwards that they appreciated the broadness of the course, in that the tools and techniques can be applied to a wide range of interventions.
The course was slightly restructured after this summer school, with more interaction brought earlier on in the course, as the first part of the course was previously quite theory-heavy.
A full-day pilot is planned for July 2026 with a smaller group made up largely of Fellows from the Software Sustainability Institute Fellowship scheme; this smaller group has more experience in institutional change (though likely no training on the topic) and is a group we know well, so are well placed to be \enquote{critical friends} and give honest, constructive feedback.

Both pilot groups were drawn from already-engaged populations -- the NetDRIVE summer school attracts PhD students already interested in Net Zero DRI, and the SSI Fellows are individuals already committed to open and sustainable research practice. Future work will focus on identifying and piloting the course with less pre-committed audiences, to test whether the material lands as effectively with participants who have not already self-selected into sustainability-focused communities.

A standardised evaluation instrument has been developed and published as part of the course resources (see \url{https://kirstypringle.github.io/advocacy-for-sustainable-research/resources/resources/\#course-evaluation}), so that others delivering the full course can evaluate it consistently. The instrument is a post-course survey covering perceived usefulness and appropriateness of content (amount and level), participants' prior familiarity with advocacy and organisational change, which parts of the course were most useful, what did not work well, whether the course changed how participants think about advocacy, and whether participants intend to act differently as a result. The pilots described above used a lighter-touch, informal approach to feedback rather than this instrument, as is appropriate for early-stage piloting; the standardised instrument is intended for use once the course is delivered more widely.

\section{Conclusion}
Many existing sustainability training resources focus on understanding environmental impacts, measuring emissions, or identifying technical interventions. These are important foundations, but there is a frequent assumption that raising awareness and building evidence will naturally lead to organisational change. Unfortunately, the evidence presented in this paper's Section \ref{background} suggests that this assumption may not hold. Researchers and university staff are frequently concerned about environmental sustainability and they work at institutions that already possess sustainability policies and strategic commitments. However, translating this interest and these policy commitments into changes in practice remains challenging.

The course addresses this gap by teaching advocacy and organisational change skills directly. It signposts existing resources to give foundational knowledge of environmental impacts, as well as including a background section on emissions and the footprint of digital research for those who need it. But the core content focuses on how to create change rather than why change is needed. Although designed for people working in research environments, the skills and concepts taught are broadly applicable.

The course is designed as a dual-layer resource: slides support facilitated delivery in a one-day workshop setting, while extended background notes make the same material accessible for self-directed study. Both layers are generated from the same open-source markdown files and published online. The course is published under a CC BY 4.0 licence and we welcome feedback, contributions, and collaboration from the community - particularly from those who have delivered similar content or who wish to adapt the material for their own context. The material remains at an early stage, with one pilot completed and a second planned for July 2026, but we believe this course fills a genuine gap in the training landscape for researchers who want to drive environmental change in their institutions.

\bibliographystyle{unsrtnat}
\bibliography{references}

@book{kotter1996leading,
  author    = {Kotter, John P.},
  title     = {Leading Change},
  publisher = {Harvard Business School Press},
  address   = {Boston, MA},
  year      = {1996}
}

@book{buganza2023storymaking,
  author    = {Buganza, T. and Bellis, P. and Magnanini, S. and Press, J. and Shani, A.B. and Trabucchi, D. and Verganti, R. and Zasa, F.P.},
  title     = {Storymaking and Organizational Transformation: How the Co-creation of Narratives Engages People for Innovation and Transformation},
  year      = {2023},
  publisher = {Routledge}
}

@book{denning2000springboard,
  author    = {Denning, S.},
  title     = {The Springboard: How Storytelling Ignites Action in Knowledge-era Organizations},
  year      = {2000},
  publisher = {Butterworth and Heinemann},
  address   = {Boston}
}

@misc{pringle2026advocacy,
  title     = {Advocacy for Sustainable Research},
  author    = {Pringle, Kirsty and Smith, Lorna and Ochu, Erinma and Wilson, Greg},
  year      = {2026},
  url       = {https://kirstypringle.github.io/advocacy-for-sustainable-research/},
  note      = {Licensed under CC BY 4.0}
}

@article{parkermoon2023digital,
  author  = {Parker Moon, Z. and Palmerini, P. and Drayton, J. and Noon, R. and Gibson, K. and Gold, L. and Ochu, E.},
  title   = {Digital storytelling: A relational pedagogic approach to rebuilding hybrid places for creativity, equity and community building in a crisis},
  journal = {Advances in Online Education},
  year    = {2023},
  volume  = {2},
  pages   = {1--18}
}

@techreport{snow2021storytelling,
  author      = {Snow, T. and Murikumthara, D. and Dusseldorp, T. and Fyfe, R. and Wolff, L. and McCracken, J.},
  title       = {Storytelling for Systems Change: Insights from the Field},
  year        = {2021},
  month       = {November},
  institution = {Centre for Public Impact; Dusseldorp Forum; Hands Up Mallee},
  url         = {https://centreforpublicimpact.org/wp-content/uploads/2024/10/storytelling-for-systems-change-report.pdf},
  note        = {Last accessed 26 June 2026}
}

@misc{concordat2024,
author       = {{Wellcome}},
title        = {Concordat for the Environmental Sustainability of Research and Innovation Practice},
year         = {2024},
url          = {https://wellcome.org/about-us/positions-and-statements/environmental-sustainability-concordat},
note         = {Accessed 2026-06-19}
}

@misc{ukri2025,
author       = {{UK Research and Innovation}},
title        = {UKRI Environmental Sustainability Strategy 2025--2030},
year         = {2025},
url          = {https://www.ukri.org/publications/ukri-environmental-sustainability-strategy/ukri-environmental-sustainability-strategy-2025-to-2030/},
note         = {Accessed 2026-06-19}
}

@article{dablander2024,
author  = {Dablander, Fabian and Sachisthal, Maien S. M. and Cologna, Viktoria and others},
title   = {Climate Change Engagement of Scientists},
journal = {Nature Climate Change},
year    = {2024},
volume  = {14},
pages   = {1033--1039},
doi     = {10.1038/s41558-024-02091-2}
}

@misc{oxford2026,
author       = {{University of Oxford}},
title        = {Environmental Sustainability Survey 2025: Key Findings},
year         = {2026},
url          = {https://sustainability.admin.ox.ac.uk/article/environmental-sustainability-survey-2025-key-findings},
note         = {Accessed 2026-06-20}
}

@techreport{juckes2023,
  author       = {Juckes, Martin and Bane, Michael and Bulpett, James and 
                  Cartmell, Kate and MacFarlane, Mark and MacRae, Malcolm and 
                  Owen, Andrew and Pascoe, Charlotte and Townsend, Philip},
  title        = {Sustainability in Digital Research Infrastructure: 
                  UKRI Net Zero DRI Scoping Project Final Technical Report},
  year         = {2023},
  doi          = {10.5281/zenodo.8199984},
  url          = {https://zenodo.org/records/8199984},
  note         = {Accessed 2026-06-20}
}

\end{document}